\documentstyle[12pt]{article}
\begin{document}
{\bf \noindent Preprint Bern BUHE-94-05\\
submitted to Phys. Rev. D, August 1994\\}
\begin{center}
{\large \bf
Detection of nuclear recoils in prototype dark matter detectors, made
from Al, Sn and Zn Superheated Superconducting Granules}\\
\vspace{1.5cm}
M. Abplanalp, C. Berger, G. Czapek, U.~Diggelmann, M. Furlan,
A.~Gabutti$^a$, S.~Janos, U.~Moser, R. Pozzi, K. Pretzl, K. Schmiemann\\
{\it Laboratory for High Energy Physics,  University of Bern, Sidlerstrasse 5,
CH 3012 Bern, Switzerland}\\
D. Perret-Gallix\\
{\it LAPP, Chemin de Bellevue, 74941 Annecy, France}\\
B. van den Brandt, J.A. Konter, S. Mango\\
{\it Paul Scherrer Institute, CH-5232 Villigen PSI, Switzerland}
\end{center}
\begin{center}
\section*{Abstract}
\end{center}
This work is part of an ongoing project to develop a Superheated
Superconducting Granule
(SSG) detector for cold dark matter and neutrinos.
The response of SSG devices to
nuclear recoils has been explored irradiating SSG detectors with a 70MeV
neutron beam. The aim of the experiment was to test the sensitivity of Sn,
Al and Zn SSG detectors to nuclear recoil energies down to a few keV.
The detector consisted of a hollow teflon cylinder (0.1cm$^3$ inner volume)
filled with tiny superconducting metastable granules embedded in a
dielectric medium.
The nuclear recoil energies deposited in the SSG were determined
measuring the neutron scattering angles with a neutron hodoscope.
Coincidences in time between the SSG and the hodoscope signals have been
clearly established. In this paper the results of the neutron irradiation
experiments at different SSG intrinsic thresholds are discussed
and compared to Monte Carlo simulations. The results show that SSG are
sensitive to recoil energies down to $\sim$1keV. The limited angular resolution
of the neutron hodoscope prevented us from measuring the SSG sensitivity to
even lower recoil energies.\\
PACS numbers: 95.30.C, 07.62
\\[6ex]
$^a$ present address:\\
Max-Planck-Institut f\"ur Physik, D-8000 M\"unchen 40, Germany

\newpage

\section{Introduction}
Superheated Superconducting Granule (SSG) detectors are presently being
developed for the detection of neutrinos and cold dark matter candidates
\cite{ref1}.
Various astronomical observations on the dynamics of spiral galaxies and galaxy
clusters seem to suggest that most of the matter ($\sim$90\%)
in the universe is in the form of dark matter.
Models of large structure formations combined with the COBE discovery of
anisotropies in the cosmic microwave background predict a dark matter
scenario made of a mixture of relativistic (hot) and non relativistic (cold)
particles.
Possible cold dark matter candidates are weakly interacting massive particles
(WIMPs) with masses between 30GeV and some
TeV. A recent review on dark matter can be found in Ref. \cite{ref2}.
After the proposal of Drukier and Stodolsky \cite{ref:Stodolsky} and the
work of Goodman and Witten \cite{ref:Goodman} attention has been devoted to
the possibility of a direct detection of WIMPs via elastic neutral current
scattering with nuclei.
Assuming a dark matter halo gravitationally trapped in the galaxy,
appreciable counting rates can be reached with recoil energy thresholds
of a few keV \cite{ref:Gabutti}.

The interest in superconducting devices for dark matter detection is based on
the very small quantum energies involved to break a Cooper pair
($\simeq$ 1meV) compared to conventional ionization ($\simeq$ 20eV)
and semiconductor detectors ($\simeq$ 1eV).
Superheated Superconducting Granule (SSG) detectors  \cite{ref:Pretzl}
consist of a collection
of tiny (some $\mu$m diameter) spheres embedded in a dielectric
medium. The granules, made of a type I superconductor having a
superheating and supercooling phase transition, are kept at a
constant temperature in the metastable state.
The detection principle is based on the phase transition from the metastable
to the normal conducting state (flip) of a single granule due to
the energy deposited by the interacting particle.
The detection threshold is given by the minimum energy  needed to raise the
granule temperature from the operating to the transition temperature. Due to
the disappearance of the Meissner effect, the magnetic field enters the normal
conducting granule, and therefore
the phase transition can be sensed with a pickup loop around the detector.
In principle, very low energy thresholds ($\sim$100eV) can be obtained with
an appropriate choice of the granule size, the superconducting
material, the operating temperature and the
strength of the applied magnetic field.

The sensitivity of SSG detectors to minimum ionizing particles
\cite{ref:Mips}, x-rays \cite{ref:Xrays} and $\alpha$ particles
\cite{ref:Frank} has been proven in the past.
To study the response of the detector to nuclear recoil
energies in the keV range, a set of experiments has been performed by our
group  irradiating Sn, Al and Zn SSG detectors with a 70MeV neutron beam
at the Paul Scherrer Institute in Villigen (Switzerland).
The investigated range of nuclear recoil energies was from 1keV up to some
hundred keV
and is in turn comparable with the recoil energies
expected in WIMP-nucleus interactions.
Due to the fast transition time of the granules ($\sim$100ns \cite{ref:Miha}),
coincidences in time between the SSG signals and the neutron beam were
clearly established, making irradiation tests a powerful probe to
evaluate the response to small recoil energy depositions.

In this paper, the results of the irradiation experiments are discussed
and compared to Monte Carlo simulations. In the investigated range of
nuclear recoil energies, we have found good agreement between the
performance of Al SSGs  at different detector thresholds and the
theoretical expectations. In the case of Sn SSG detectors, the measured
response to nuclear recoils (sensitivity) was found to be
higher than expected. The sensitivity of Zn SSGs was found to be correlated to
the time difference between the scattering interaction and the onset of the
resulting phase transition.
Energy thresholds down to $\sim$1keV were reached in  Sn and Zn SSG
detectors.
Due to the limited angular resolution of the neutron hodoscope we were not
able to test the SSG sensitivity to lower recoil energies.
The presented results are encouraging for possible applications of SSGs for a
dark matter detector.

\section{Experimental Setup}
We performed neutron irradiation experiments on  SSG detectors made of
Al, Sn and Zn granules with diameters
of 20-25$\mu$m, 15-20$\mu$m and 28-30$\mu$m, respectively. The
setup used in the experiments is sketched in \mbox{Fig.\ref{fig:Setup}} and
the characteristics of the different SSG detectors are listed in
\mbox{Table \ref{tab2}}.

The SSG detectors consisted of a hollow teflon cylinder of 4mm inner diameter
and 8mm length, filled with a collection of granules selected in size by
ultrasonic sieving.
A scanning electron microscope analysis of a sample of sieved
granules showed a rather narrow distribution (\mbox{Table \ref{tab2}}).
The Al granules were embedded in an Al$_2$O$_3$ powder,
the Sn and Zn granules were embedded in plasticine. The volume filling
factors were between 6\% and 11\%.
The detectors were operated in the mixing chamber of a $^3$He-$^4$He dilution
refrigerator \cite{ref:Mango}. The measurements on the Zn and Sn SSGs were
done at a  temperature of 40$\pm$5mK using a pair of identical
SSG detectors mounted one behind the other along the beam axis in order to
increase the statistics. The irradiation tests with Al SSGs
were performed at 120$\pm$10mK and are described in Ref. \cite{ref:Nimneutron}.

Each SSG target was surrounded by one pickup coil (180-228 windings,
depending on
the sample) connected to a J-FET preamplifier working at room temperature.
The signal due to the phase
transition  of a single granule was a
damped oscillation with a period of about 1.5$\mu$s as shown in
\mbox{Fig \ref{fig:Signal}}.
The signal was well above the electronic noise, and  we were able to
clearly recognize the phase transition of a single granule in SSG
detectors made of more than 1 million granules.
The magnetic field was generated by a Helmholtz coil outside the cryostat
and the uncertainty in the absolute field strength was evaluated to be 1\%.

The neutron beam was produced by irradiating a Be target with 72MeV protons
from the Injector I at the Paul Scherrer Institute \cite{ref:Nbeam}.
The residual protons were swept away by a bending magnet downstream of
the Be target. The neutron
beam was collimated to a diameter of 3.2mm by a copper
collimator of 150cm length aligned with the SSG detectors.
The energy distribution of the neutrons was
sharply peaked at 70MeV with a small shoulder towards lower energies
\cite{ref:Nbeam}. The beam repetition rate was 17MHz with a bunch length of
2ns.

The scattered neutrons were detected with a counter hodoscope located 2m
behind the SSG. The neutron hodoscope consisted of 18 scintillator bars
with a length of  150cm and a cross section of 5x5cm$^2$.
The bars were arranged in three successive planes of six elements each.
The position of the interacting neutron inside a single element was determined
measuring the time difference
between the signals of two photomultipliers located at the ends of the bar with
a resulting spatial resolution of $\pm$4cm corresponding to an angular
resolution of $\pm$20mrad.
The hodoscope covered scattering angles from 20mrad up to about 0.45rad.

In the earlier Al runs \cite{ref:Nimneutron}, the neutron beam was monitored
using a 5cm thick scintillation counter behind the hodoscope preventing an
absolute evaluation of the beam intensity. To discriminate against charged
particles, two veto counters were installed between the
cryostat and the hodoscope. The first counter was a 5mm thick scintillator
located 20cm behind the SSG and covering an area of 20x20cm$^2$. The energy
threshold was set to 0.15MeV. The second veto counter was a 2cm thick
scintillator located close to the hodoscope covering the full
acceptance.

In the Zn and Sn irradiation tests, we improved the setup using concrete
blocks covering the sides and the top of the neutron
hodoscope to shield against
background from the surrounding. In addition, the neutron flux was
measured using a thin CH$_2$ target positioned after the collimator and a
telescope at 13 degrees to detect protons coming from $n-p$ reactions in this
target.  To discriminate against charged particles entering the SSG detector
an additional 5mm scintillator veto counter was located in front of the
cryostat window.

\section{Irradiation experiments}
\label{sect:Irrad}

The distribution of the superheating field of the granules within the detector
was measured without irradiation by simply ramping the
field and recording the rate of phase
transition signals as a function of the applied field. The superheating field
distributions of the SSG detectors used in the irradiation experiments
are shown in \mbox{Fig.\ref{fig:Sh}}.
Previous works on single granules \cite{ref:Frank} have shown a
modulation in the superheating field related to the orientation of
the granule with respect to the magnetic field. The crystalline structure of
the granules, deviations from spherical shape and magnetic
field distortions across the granules were found to be
responsible for the observed  superheating field distributions.

\subsection{Irradiation cycles}

The irradiation experiments consisted of many consecutive cycles. In each
cycle, the magnetic field was changed with time as shown in
\mbox{Fig.\ref{fig:Cycle}}. The detector
threshold was set ramping the magnetic field up to a reference value $B_1$.
Then the field was reduced to a lower value $B_2$ and maintained constant
for 5 or 10 minutes during which the beam induced phase transitions were
recorded. Finally, the field was raised to 50mT in order to flip
the remaining superconducting granules into the normal state.
The detector magnetic threshold was defined as $h=1-B_{2}/B_{1}$, the
values of the reference field $B_1$ being listed in \mbox{Tab.\ref{tab2}}.
Due to the superheating field distribution only granules with an individual
field $B_{sh}>B_{1}$ were sensitive to the neutron beam.
Data were taken at detector thresholds $h$ in the range from 0.5\% up to 10\%,
which correspond to recoil energy thresholds (see chapter \ref{sect:Eth})
as shown in \mbox{Fig.\ref{fig:Eth}}.
To align the SSG detectors to the beam, the cryostat was moved perpendicular
to the beam axis to find the maximum SSG counting rate
without selecting coincidences with the hodoscope.
The measured beam profile was a gaussian with a standard
deviation of $\sigma \sim$2mm as it is shown in \mbox{Fig.\ref{fig:Beamprof}}.
The fraction of the total neutron flux impinging on the SSG detector
was 35$\pm$5\%.

The decrease of the counting rate within an irradiation cycle is
shown in
\mbox{Fig.\ref{fig:Decay}} for a Sn SSG detector at two magnetic thresholds.
This is due to the loss of sensitive granules during
irradiation.  At high detector thresholds, the decay is
less pronounced because only a small fraction of the granules is sensitive to
the impinging particles.

Due to the small neutron cross section and the small size of the target,
the probability of having multiple elastic scatterings within the SSG is
rather small. Neutrons mainly  produce the phase transition of a single
granule. Charged particles instead can cause more than one granule to flip.
Single flip events were clearly distinguished
from events with higher multiplicity using the pulse height
of the phase transitions signals of the SSG detectors
(\mbox{Fig.\ref{fig:multe}}).

\subsection{Selection cuts}

In order to select a neutron induced event, coincidences in time between the
injector radio frequency, the SSG detector and the hodoscope signals had to be
established. The SSG trigger was defined by the first zero crossing
of the SSG signal (\mbox{Fig.\ref{fig:Signal}}).

Typical time distributions between the SSG and hodoscope signals
are shown in \mbox{Fig.\ref{fig:Digi}} for the Al, Zn and Sn SSG detectors.
The SSG signals in coincidence were well above the accidental background and
only events within the hatched regions were selected for the analysis.
The events outside the time window were used to evaluate the accidental
background, which was subtracted from the selected events after normalization.
The standard deviations from the means of the distributions of the SSG signals
in coincidence above background were about
25ns in Sn, 45ns in Al and 150ns in Zn.

In contrast to Sn, the coincidences in the Zn irradiation test lie within
a wide range of time differences ($\simeq$800ns)
between the hodoscope trigger and the
SSG flip signals, as shown in \mbox{Fig.\ref{fig:Digi}}.
Recent investigations \cite{ref:toledo} have shown that the observed time
distributions are not affected by the speed of the magnetic field penetration
inside the granule, but are due to the thermal relaxation of the quasiparticles
and phonons produced by the recoil event. The time scale of the thermal
diffusion is shorter in Sn than in Zn due to the shorter quasiparticle
relaxation times \cite{ref:Kaplan}.

To reduce the background from charge exchange reactions, which produced a
fast proton in the final state, only events with
no signal in the veto counters were considered.
The kinetic energy of the interacting neutron was determined from the time
of flight between the neutron production target and the
hodoscope. A cut on the
scattered neutron time of flight of $\pm$1.5ns was introduced as shown in
\mbox{Fig.\ref{fig:Tof}}.
Due to the
bunch width of the beam, the accuracy on the energy measurement of a 70MeV
neutron  was $\pm$8MeV.

After the cuts, the number of events used in the final analysis
was approximately 300 in Al and 1500 in Zn and Sn for each detector threshold.

\section{Simulation of the experiment}
\label{sect:Sim}

To evaluate the SSG response to nuclear recoils, the
measurements were compared to a Monte Carlo simulation of the experiment.
The SSG response to the beam was calculated considering only elastically
scattered neutrons.

\subsection{Nuclear recoil energies}
\label{sect:Nucl}

The
distribution of the scattering angles ($\vartheta$) was evaluated by a partial
wave expansion using the optical model \cite{ref:Optmod}. This
approximation can be safely used because the de Broglie wave length of a
70MeV neutron (0.54fm) is smaller than
the nuclear radius $R=1.37A^{1/3}$fm with $A$ the atomic number.
Because the neutron kinetic energy of 70MeV is higher than the nuclear
potential of about 30MeV \cite{ref:nucpot}, the neutron wave enters the nucleus
with a propagation vector shifted by k$_1 \simeq$ 0.3 fm$^{-1}$ and is absorbed
by neutron-nucleon-scattering with an absorption coefficient of
K=(0.25-0.29)fm$^{-1}$, depending on the nuclear neutron-proton-ratio
\cite{ref:Optmod}. The differential elastic cross section is given by:
\begin{equation}
\frac{d\sigma_{el}}{d\vartheta} = \frac{\pi}{2 k^2} \mid f(\vartheta) \mid
^2 sin(\vartheta)
\label{eq:Scatang}
\end{equation}
\begin{eqnarray*}
f(\vartheta) = \sum_{l=0}^{l+1/2<kR} (2l+1)\,[1-exp(-K+2\,i\,k_1)\,sl]
\,P_l(\cos\vartheta) \\
sl = 1/k\,[\,(kR)^2 - (l+1/2)^2\,]^{1/2}
\end{eqnarray*}
with $k$ the de Broglie propagation vector of the neutron and
$P_l$ the Legendre polynome of degree $l$. Due to nuclear absorption, the
total elastic cross section differs from the geometrical cross section $\pi
R^2$, depending on the value of $K R$, leading to total cross sections at 70MeV
of 0.82barn, 1.39barn and 1.85barn for Al, Zn and Sn, respectively
\cite{ref:Optmod}.
The scattering probabilities listed in  \mbox{Tab.\ref{tab2}} are calculated
considering an effective target thickness d of 0.5mm, 1.9mm and 0.9mm for the
Al, Zn and Sn SSG, respectively.
The recoil energy $E_r$ deposited inside a granule by an elastically
scattered neutron was evaluated from the scattering angle $\vartheta$.

\subsection{Energy threshold}
\label{sect:Eth}

The energy threshold E$_{th}$ for a single granule is given by
the energy needed to raise the granule temperature from the bath temperature
T$_b$ to the transition temperature T$_{sh}$.
{}From the phase diagram, this change in temperature can be related to the
magnetic threshold $h_{g}=1-B_{2}/B_{sh}$ with $B_{2}$ and $B_{sh}$ the
applied and the superheating field respectively.
The transition temperature $T_{sh}$, evaluated from the approximated
temperature dependence of the superheating field $B_{sh}=B_{o}(1-t^{2})$ is:
\begin{equation}
T_{sh}(h_g) = T_{c} \sqrt{t^2+h_{g}(1-t^2)}
\label{eq:Tsh}
\end{equation}
with $T_c$ the critical temperature at
zero magnetic field and $t=T_{b}/T_{c}$ the reduced temperature of the bath.

For a granule of volume V, the energy density threshold $q_{th}=E_{th}/V$
required to produce a phase transition can be defined as:
\begin{equation}
q_{th} = \int \limits_{T_b}^{T_{sh}(h_g)} C(T)\,dT
\label{eq:q}
\end{equation}
where the superconducting specific heat $C(T)$ is given by:
\begin{equation}
C(T) = \beta \,\frac{T^3}{\theta ^3} + \gamma \, T_c \, u(t)
\end{equation}
using the parameterization \cite{ref:specificheat}:
\begin{eqnarray*}
\begin{tabular}{lcl}
{\it u(t) =} & {\it 3.33 exp(-1.76/t) t$^{-3/2}$} & ,{\it 0 $<$t$<$0.1} \\
{\it u(t) =} & {\it 26   exp(-1.62/t)}            & ,{\it 0.1$<$t$<$0.161} \\
{\it u(t) =} & {\it 8.5 exp(-1.44/t) }            & ,{\it 0.161$<$t$<$1} \\
\end{tabular}
\end{eqnarray*}
and $\beta$=1944 $J mol^{-1} K^{-1}$. The values of the  Sommerfeld
constant $\gamma$, the Debye temperature $\theta$ and the critical
temperature $T_c$ are listed in \mbox{Tab.\ref{tab1}} for the three materials
of interest \cite{ref:Handb}.

In eq. \ref{eq:q}, the energy deposited
by the interacting particle is assumed to be uniformly distributed over the
whole granule.
Such a scenario is usually referred to as {\it global heating model}
\cite{ref:Pretzl}.
The calculated energy thresholds for 25$\mu$m Al, 29$\mu$m Zn
and 17$\mu$m Sn granules are shown in \mbox{Fig.\ref{fig:Eth}}.
The recoil energies due to neutral-current scattering of dark matter are
expected to be of the order of a few keV \cite{ref:Gabutti}.
Note that such energy
thresholds can be reached with the presently used superconducting granules and
in principle even lower ones could be reached using smaller granules.

In the irradiation
experiments, the detector threshold $h$ was set by
ramping the magnetic field as
shown in \mbox{Fig.\ref{fig:Cycle}} and discussed in section
\ref{sect:Irrad}. Only granules with an individual field $B_{sh}\geq B_1$
were sensitive.
The probability of having a phase transition in the Sn SSG
for a given energy deposition, shown in \mbox{Fig.\ref{fig:Flipprob}},
can be evaluated from \mbox{eq.\ref{eq:q}} using the measured granule size and
the superheating field distributions.
The slow increase of the flip probability with increasing recoil energy is due
to the superheating field distribution.

\subsection{Occurrence of neutron induced phase transitions}

In the simulation program, a phase transition was triggered depending on the
energy deposit by an elastically scattered neutron inside a granule. This
granule was selected from a pool of granules with a flat size distribution
within the range given in \mbox{Tab.\ref{tab2}} and a superheating field
distribution obtained from the measurements.
The distribution of scattering angles was derived from
\mbox{eq.\ref{eq:Scatang}}. A phase transition occurred whenever the deposited
energy density q=E$_r$/V exceeded the threshold value q$_{th}$
(\mbox{eq.\ref{eq:q}}).
All the flipped granules ($q\geq q_{th}$) were made insensitive to later
interactions.
After each phase transition, the path of the scattered neutron was tracked
to the hodoscope.
Events with no hits in the hodoscope were discarded.
The scattering angles and recoil energies associated to the selected events
were stored. For each of these events the hit pattern of the hodoscope was
evaluated. The spatial resolution of the neutron hodoscope was
properly taken into account.
During the irradiation, the probability of having a phase transition inside the
SSG detector for a given recoil energy changes with time because of the
decrease in the number of sensitive granules (\mbox{Fig.\ref{fig:Decay}}).
To account for this effect, the Monte Carlo program generated the same number
of phase transitions as measured in a typical cycle for each detector
threshold.
The total number of scattering events generated in the simulations
was used to normalize the calculated distributions at different detector
thresholds.

\section{Experimental results}
\label{sec:expresults}
\subsection{Normalization}
\label{sec:norm}
To evaluate the absolute SSG efficiency to nuclear recoils, the rate of
coincidences was normalized to the beam monitor counts and divided by the
hodoscope counting rate. To estimate the hodoscope counting rate due to
neutrons scattered by the SSGs, additional runs with bulk material were
performed using
a thick Sn target replacing the SSG detector. Elastically scattered
neutrons were selected requiring a trigger in the hodoscope with the same
selection cuts on the neutron time of flight, on the hodoscope and on the veto
counters as used in the SSG irradiation runs. The bulk
runs were done with two different target thicknesses (2mm and 10mm). After
subtraction, the resulting hodoscope counting rate for an effective 8mm thick
Sn bulk was about 3.2$\pm$0.3(stat) events per beam monitor count. The expected
hodoscope counting rate n$_{Sn}$ due to elastic scatterings from the Sn SSG
detectors was evaluated from the bulk runs considering the ratio between the
effective thicknesses of the SSG and the bulk target. Since the neutron flux
impinging the SSG detectors was only 35$\pm$5\% of the total beam flux, the
hodoscope counting rate n$_{Sn}$ normalized to the beam monitor was evaluated
to be n$_{Sn}$=0.13$\pm$0.02. The error contains the statistical and
systematical errors of the bulk runs and the uncertainty in the beam coverage
of the SSG detectors. The Sn bulk measurements were also used to evaluate the
hodoscope counting rate n$_{Zn}$ due to neutrons elastically scattered from the
Zn SSG detectors. The probability S for an incoming neutron to be
elastically scattered is higher in the Zn than in the Sn detectors (see
table \mbox{\ref{tab2}}). From the Monte Carlo simulations,
the mean geometrical acceptance of the hodoscope for
elastically scattered neutrons on Zn nuclei turned out to be
about 10\% higher than on Sn nuclei. The hodoscope
counting rate is with n$_{Zn}$=0.39$\pm$0.06 about three times higher
than n$_{Sn}$.

In the irradiation runs, the rates of coincidences between the SSG detectors
and
the neutron hodoscope have been measured. The loss of sensitive granules during
irradiation (\mbox{Fig.\ref{fig:Decay}}) was taken into account. The ratio
between the SSG coincidence rate and the hodoscope counting rate
defines the detection efficiency. It reflects the
ability of the SSG detector to detect events with a scattered
neutron in the hodoscope. The efficiency will never reach 100\% since a
fraction of the granules is never sensitive due to the width of the
superheating curve and the choice of the magnetic threshold.

\subsection{Aluminum}
The recoil energies of the selected events for three detector thresholds are
compared with the Monte Carlo results using the global heating model in
\mbox{Fig.\ref{fig:erecal}}.

The experimental uncertainty in the determination of the recoil energy is due
to the neutron kinetic energy resolution of the time of flight measurement and
the scattering angle resolution of the hodoscope. The uncertainty in the
counting rate is due to statistical and systematic errors in the selection
cuts. In the full range of applied detector thresholds,
the experimental data agree well with the global heating model.
The distributions shift to
higher values when the detector threshold is increased. These shifts are
not clearly seen due to the different magnetic thresholds of the sensitive
granules and to the limited recoil energy resolution which corresponds to the
bin width in \mbox{Fig.\ref{fig:erecal}}

The agreement between experimental data and Monte Carlo results
also appears in the comparison of the Monte Carlo detector efficiency
(for scattering events with the final neutron being detected in the hodoscope)
and the experimental efficiency, as shown in \mbox{Fig. \ref{fig:effal}}.
The
hodoscope counting rate n$_{Al}$ in this case was determined by a calibration
run with a 4cm thick Al bulk in the beam.
Its thickness was chosen so, that
most of the neutrons in the beam are scattered within the bulk. The counting
rates were normalized to a 5cm thick scintillation counter placed in the beam
behind the hodoscope.

As expected, the measured detector efficiency decreases
with increasing
detector threshold. Due to the distribution of recoil energies and the spread
of
superheating fields, not every elastic scattering releases enough energy to
flip the granule, so that even at zero detector threshold the efficiency
does not reach 100\%

The experimental results show that the behaviour of Al
SSGs is consistent with global heating.

\subsection{Zinc}
The Zn data were divided into two sets selecting events with either short
(1.2-1.6$\mu$s) or long (1.6-2.0$\mu$s) time differences between the hodoscope
and the SSG signals. In \mbox{Fig.\ref{fig:ereczn}}, the recoil energy
distributions of the two sets of data
are compared to the global heating Monte Carlo simulation.
For clearness, the Monte Carlo and the data have been normalized
to the same total number of events.
\mbox{Fig.\ref{fig:ereczn}} exhibits a clear correspondence between a fast
(slow) response to a large (small) recoil energy.

In \mbox{Fig. \ref{fig:effzn}} the measured efficiency considering both slow
and fast triggers is compared to the Monte Carlo simulations.
The Monte Carlo results agree in general well with the experiment, but the
decrease of efficiency with increasing detector threshold is steeper in
the experimental data
than in the Monte Carlo simulations. This disagreement still remains when only
the slow signals are considered.

\subsection{Tin}
\label{sec:tin}
Previous experiments on Sn SSG irradiated with $\alpha$ particles
\cite{ref:Frank} and minimum ionizing particles \cite{ref:Mips} have shown that
the detector sensitivity is higher than expected by the global heating model
(eq. \ref{eq:q}). To evaluate the SSG sensitivity to nuclear recoils, the
measurements were compared to the Monte Carlo simulations using
the global heating model.

The comparison between the measured and the
calculated recoil
energy distribution is shown in \mbox{Fig.\ref{fig:erecsn}} for four detector
thresholds. The observed sensitivity of Sn SSG to
nuclear recoils of some keV demonstrates the ability of such devices to detect
also nuclear recoils due to WIMP interactions.
The data seem to indicate that Sn granules are more sensitive to
nuclear recoils than expected using the global heating model.

\mbox{Fig. \ref{fig:effsn}} compares the measured efficiencies
to detect a neutron scattering with the final state neutron being registered
in the hodoscope
at six different detector thresholds with the predictions from the Monte Carlo
algorithm using the global heating
model.

The higher sensitivity of Sn SSGs is still under investigation. It may be
explained by a heat diffusion effect which takes into account the location of
the nuclear recoil inside the granule together with the minimum size of the
nucleation center at the surface of the granule \cite{ref:Albiphysrev}.

\section{Conclusions}

Irradiation measurements of Al, Zn and Sn superheated superconducting
granule (SSG) detectors exposed to a 70MeV neutron beam have been performed
to study
the sensitivity to nuclear recoils of defined energy down to some keV. The
limited angular resolution of the experiment prevented us to measure the
sensitivity of the SSG to recoil energies below 1keV. In the case of Al SSGs,
the measured recoil energies follow the theoretical predictions of
the global heating model, while the Sn SSGs show a higher sensitivity to
nuclear recoils than predicted by this model. In the case of
Zn SSGs a large time difference between the interaction and the occurrence
of the flip signal has been observed. This delay could be attributed to the
heat diffusion mechanism,
which reflects the long quasiparticle relaxation time in Zn.

The experiments confirmed that SSG detectors can be successfully
used to detect recoil
energies even below 1keV, depending on the material and the granule size. By
improving the readout circuit in order to detect signals from smaller granules,
even smaller energy thresholds (in the eV region) could be reached.
The detector efficiency does not saturate above the energy threshold because of
the
shape of the superheating curve and the size distribution of the granules. The
obtained sensitivities encourage the use of SSG detectors for WIMP searches.
\section{Acknowledgments}
We would like to thank K.~Borer, M.~Hess, S.~Lehmann, L.~Martinez, F.~Nydegger,
J.C.~Roulin and H.U.~Sch\"utz from the High Energy Physics Laboratory of the
University of Bern for technical support. We would also like to thank
P.A. Schmelzbach, the coordinator of the Philips Cyclotron at the Paul
Scherrer Institute. This work was supported by the
Schweizerischer Nationalfonds zur F\"orderung der wissenschaftlichen Forschung
and by the Bernische Studienstiftung zur F\"orderung der wissenschaftlichen
Forschung an der Universit\"at Bern.


\newpage
\section*{Tables}

\begin{table}[h]
\begin{center}
\caption{\label{tab2} SSG detector characteristics.}
\begin{tabular}{c||c|c|c|c|c}\hline
mate- & B$_{1}$ [mT]  & $T_b$ & grain dia-  & Filling     & scattering \\
rial  &               & [mK]  & meter [$\mu$m] & factor [\%] & prob. S[\%]
\\\hline\hline
Al       &  10.4    & 120 & 20-25   &    6  & 0.24 \\
Zn       &  6.5     & 40  & 28-30   &   11  & 1.73 \\
Sn       &  31.5    & 40  & 15-20   &    6  & 0.60 \\\hline
\end{tabular}
\end{center}
\end{table}
\begin{table}[bh]
\begin{center}
\caption{\label{tab1} Properties of the used SSG materials.}
\begin{tabular}{c||c|c|c}\hline
         & T$_c$  & Sommerfeld & Debye Temp. \\
material & [K]    & constant   & [K] \\
         &        & [mJ/mol/K$^2$]&  \\\hline\hline
Al   & 1.2  &  1.36 & 426 \\
Zn   & 0.9  &  0.63 & 310 \\
Sn   & 3.7  &  1.75 & 195 \\\hline
\end{tabular}
\end{center}
\end{table}


\section*{Figure Captions}
\newcounter{fig}
\begin{list}{Fig. \arabic{fig}:}{\usecounter{fig}}

\item\label{fig:Setup} Experimental setup.

\item\label{fig:Signal}
Signal output of the readout electronics for a beam induced  phase
transition of a single granule in a Zn detector.

\item\label{fig:Sh}
The superheating curves of the a) Al, b) Zn and c) Sn SSG detectors
as measured in the experiment.

\item\label{fig:Cycle}
Magnetic field versus time during one irradiation cycle.

\item\label{fig:Eth}
The corresponding recoil energy thresholds E$_{th}$ versus the SSG
detector thresholds $h$ for:
$a)$ Al granules with 23$\mu$m diameter at $T_b$=120mK, $b)$ Zn granules with
29$\mu$m
diameter at $T_b$=40mK, c) Sn granules with 17$\mu$m diameter at $T_b$=40mK.

\item\label{fig:Beamprof}
The neutron beam profile measured with the Sn SSG detector
across the horizontal beam axis.

\item\label{fig:Decay}
Time evolution of the Sn SSG detector counting rate
during irradiation at two magnetic thresholds, not requiring
coincidence with the hodoscope.

\item\label{fig:multe}
Signal amplitudes of grain flips in the
Sn SSG detector without requiring a coincidence with the neutron hodoscope.
Single grain flips can be clearly distinguished from multiple (2,3,4 etc)
flips.

\item\label{fig:Digi}
Distribution of the time difference between SSG and
hodoscope signals for: $a)$ Al, $b)$ Zn and $c)$ Sn SSG detectors.
The selected events are within the hatched regions. The time scale is offset by
a constant value due to electronic delays.

\item\label{fig:Tof}
Time of flight (TOF) spectrum of incident neutrons for the Sn
detector.
The selected events are within the hatched regions.

\item\label{fig:Flipprob}
Calculated phase transition probability as a function of the
deposited recoil energy inside a granule for the Sn SSG detector with the
thresholds $h$=0.5\% and $h$=3.5\%.

\item\label{fig:erecal}
Recoil energy distribution of the Al SSG for three detector
thresholds: a) 2\%, b) 5\%, c) 10\%. Line: Monte Carlo
 prediction using the global heating model, points:
data from the irradiation runs.

\item\label{fig:effal}
Al SSG detection efficiency for an event with the elastically
scattered neutron passing the hodoscope. Points: Data from
irradiation runs, line: Monte Carlo prediction using the global heating model.

\item\label{fig:ereczn}
Recoil energy distributions of the Zn SSG for two detector
thresholds, for either short (1.2-1.6$\mu$s, Figs. a and b) and long
(1.6-2.0$\mu$s, Figs. c and d) time differences between hodoscope and SSG
signals. Line: Monte Carlo prediction using the global heating model,
points: Data from the irradiation runs.

\item\label{fig:effzn}
Zn SSG detection efficiency for an event with the elastically
scattered neutron passing the hodoscope. Points: Data from
irradiation runs, line: Monte Carlo prediction using the global heating model.

\item\label{fig:erecsn}
Recoil energy distribution of the Sn SSG for four detector
thresholds: a) 0.5\%, b) 1.0\%, c) 2.5\%, d) 3.5\%.
Points: Data from irradiation runs, line: Monte Carlo
prediction using the global heating model.

\item\label{fig:effsn}
Sn SSG detection efficiency for an event with the elastically
scattered neutron passing the hodoscope. Points: Data from
irradiation runs, line: Monte Carlo prediction using the global heating model.

\end{list}

\end{document}